\documentclass[usenatbib,referee]{mn2e}
\usepackage{epsfig}
\newcommand{\lsim}{\lower 2pt \hbox{$\, \buildrel {\scriptstyle <}\over
{\scriptstyle
\sim}\,$}} 
\newcommand{\gsim}{\lower 2pt \hbox{$\, \buildrel {\scriptstyle >}\over
{\scriptstyle
\sim}\,$}} 

\bibpunct{(}{)}{;}{a}{,}{,} \title[A census of young stellar 
populations in the warm ULIRG PKS1345+12 ]{A census of young stellar 
populations in the warm ULIRG PKS1345+12} \author[J.Rodr\'iguez Zaur\'in, J.Holt,
C.N.Tadhunter and
R.M.Gonz\'alez Delgado]{J.Rodr\'iguez Zaur\'in$^{1}$\thanks{E-mail:jrzaurin@sheffield.ac.uk}, J.Holt$^{1}$, C.N Tadhunter$^{1}$ and R.M.Gonz\'alez Delgado$^{2}$\\
$^{1}$Department of physics and Astronomy, University of Sheffield,
Sheffield S3 7RH\\ $^{2}$Instituto de Astrofisica de Andalucia(CSIC),
P.O.Box 3004, 18080 Granada, Spain}
\begin{document}

\pagerange{\pageref{firstpage}--\pageref{lastpage}} \pubyear{2002}

\maketitle

\label{firstpage}

\begin{abstract}
We present a detailed investigation of the young stellar populations
(YSP) in the radio-loud ultra  luminous infrared galaxy (ULIRG)
PKS1345+12(z=0.12), based on high resolution {\it Hubble Space
Telescope} (HST) imaging and long slit spectra taken with the{\it
William Herschel Telescope\/} (WHT) on La Palma. While the images
clearly show bright knots suggestive of super star clusters(SSC),  the
spectra reveal the presence of YSP in the diffuse light across the
full extent of the halo of the merging-double nucleus system. Spectral
synthesis modelling has been used to estimate the ages of the YSP for
both the SSC and the diffuse light sampled by the spectra.  For the
SSC we find ages t$_{SSC}$$<$6 Myr with reddenings $0.2 < E(B-V) <
0.5$ and masses $10^6 < M_{SSC}^{YSP} <  10^7$ M$_{\odot}$. In the
region to  the south of the western nucleus that contains the SSC our
modelling of the spectrum of the diffuse light is also consistent with
a relatively young age for the YSP ($\sim$5~Myr), although older YSP
ages cannot be ruled out. However, in other regions of the galaxy we
find that the spectra of the diffuse light component can only be
modelled with a relatively old post-starburst YSP (0.04 -- 1.0~Gyr) or
with a disk galaxy template spectrum.  The results demonstrate the
importance of accounting for reddening in photometric studies of SSC,
and highlight  the dangers of focussing on the highest surface
brightness regions when trying to obtain a general impression of the
star formation activity in the host galaxies of ULIRGs. The case of
PKS1345+12 provides clear evidence that the star formation histories
of the YSP in ULIRGs are complex. While the SSC represent the vigorous
phase of star formation associated with the final stages of the
merger, the YSP in the diffuse light are likely to represent star
formation in one or more of the merging galaxies at an earlier stage
or prior to the start of the merger.

Intriguingly, our long-slit spectra show line splitting at the
locations  of the SSC, indicating that they are moving at up to 450 km
s$^{-1}$ with respect to the local ambient gas.  Given their
kinematics, it is plausible that the SSC have been formed either in
fast moving gas streams/tidal tails that are  falling back into the
nuclear regions as part of the merger process, or as a  consequence of
jet-induced star formation linked to the  extended, diffuse radio
emission detected in the halo of the galaxy.

\end{abstract}


\section{Introduction}

A variety of surveys over the last two decades have revealed
populations of galaxies that emit the bulk of their radiation at
infrared wavelengths
\citep{Soifer84a,Soifer84b,Houck84,Houck85,Lefloch05,Perezgonzalez05}.
Such objects are thought to be responsible for most of the star
formation activity in the distant Universe \citep{Lefloch05}, while in
the local Universe these galaxies are classified as Luminous (L$_{ir}
> 10^{11}$ L$_{\odot}$) or Ultraluminous ($L_{ir} > 10^{12}$
L$_{\odot}$) infrared galaxies (LIRGs/ULIRGs). The morphologies of
ULIRGs clearly suggest that most of them are recent or ongoing
mergers.  The presence of an AGN  in a subset of  these objects is
also well known, with some ULIRGs  classified as QSOs at optical
wavelengths. For a more complete review of the properties of
LIRGs/ULIRGs see \cite{Sanders96}.

Much attention has been paid recently to the so-called warm ULIRGs
(mid-infrared colours of 
{\it f\/}$_{25}$/{\it f}$_{60}$$\ge$ 0.2)\footnote{The quantities 
{\it f\/}$_{25}$ and {\it f\/}$_{60}$
represent the {\it IRAS \/} flux densities (non-colour corrected) in
units of Jy at 25 and 60~$\mu$m respectively}. These objects represent
the $\sim$20-25\% of the total ULIRG population discovered by the
IRAS satellite. Most of them have AGN optical spectra, very large
molecular gas masses ($M_{H_{2}} \sim 10 ^{10}$~M$_{\odot}$: Sanders et al.
1988) and are found in an advanced merger state. In fact,
\cite{Surace99} found that 75\% of the warm ULIRGs in their sample
of 12 objects were associated with single nucleus galaxies. These
properties  suggest that such objects represent a transitional stage
between cool ULIRGs and Radio Galaxies/QSOs \citep{Sanders88}: cool ULIRGs
evolve into warm ULIRGs on their way to becoming QSOs/radio galaxies.  
Indeed, models predict that the tidal forces associated with mergers
lead to large concentrations of gas and dust in the nuclear regions of
the galaxies \citep{Mihos96,Barnes96}; such concentrations may trigger
both AGN and starburst activity. However, the timing of the AGN
relative to the major starburst triggered by the merger remains
uncertain.  Although some models assume that both starburst and AGN
will occur simultaneously, when there is sufficient concentration of
gas  in the nuclear regions \citep{diMatteo05}, the idea that cool
ULIRGs evolve eventually into QSOs \citep{Sanders88} suggests that
there may be a significant delay between the starburst and visible AGN activity.

Given that the models make specific predictions about the histories
of the star formation triggered in the course of major gas-rich
mergers,  studies of the stellar populations in ULIRGs provide useful
information about the mergers and, potentially, allow us to test the
models.  Moreover, for those ULIRGs with AGN and powerful radio jets,
the ages of the young stellar populations (YSP) can be used to
establish the order of events during the merger, for example, whether
the AGN was activated before, at the same time as, or after the
merger-induced starburst \citep{Tadhunter05,Emonts06}.  Thus studies
of the YSP have the potential to help us to understand the possible
links between ULIRGs and AGN.

In this context, there have been surprisingly few studies of the YSP
in ULIRGs. Two complementary approaches have been used in the past:  one based on
photometric analysis of images of the galaxies taken at different
wavelengths, focused mainly on studing the YSP associated with the
bright knots detected in the objects
\citep{Surace98,Surace99,Surace00a,Surace00b,Wilson06}; the other
comprising spectrocopic analysis of UV/optical spectra of more
spatially extended regions
\citep{Canalizo00a,Canalizo00b,Canalizo01,Tadhunter05}. Note that
such optical studies sample the star formation in the systems
prior to the current prodigious formation activity represented
by the ULIRG that is likely to be hidden from
view at optical wavelengths. In this paper
we combine optical spectroscopy and imaging observation in
order to investigate the YSP in the warm ULIRG/Radio Galaxy PKS1345+12

PKS1345+12 is a luminous, compact radio source \citep{Evans99,Holt03}
in an elliptical-like host galaxy (e.g. Axon et al. 2000). The galaxy
has a clear double nucleus separated by $\sim$1.8 arcsec or 
4.3~kpc\footnote{$H_0= 75$ km s$^{-1}$, $q_0 = 0.0$ assumed throughout
resulting in a scale of 2.37 kpc arcsec$^{-1}$ at z = 0.122}, and the
AGN activity is associated with the western nucleus, which has 
a spectrum characteristic of narrow line radio galaxies \citep{Evans99,Holt03}.

Significant YSP have also been detected in PKS1345+12 at optical
wavelengths
\citep{Surace98, Tadhunter05}, which may have been formed as a
consequence of the merger. Moreover, evidence for
prodigious hidden star formation activity
is provided by the detection of a large far-IR excess by the {\it IRAS
\/} satellite ($L_{IR} = 1.7 \times 10^{12}$~M$_{\odot}$, (Evans et
al. 1999)). In addition, the CO observations indicate the presence of
a substantial reservoir ($3.3 \times 10^{10}$~M$_{\odot}$) of molecular gas
within $\sim$2.5~kpc of the active (western) nucleus ($\rho >$ 2000
M$_{\odot}$ pc$^{-3}$: Evans et al. 1999), consistent with measurements in
other ultraluminous infrared galaxies \citep{Bryant99}. The double
nucleus, the distorted large-scale morphology, the presence of a rich
inter-stellar medium (ISM) and YSPs clearly indicate that PKS1345+12
represents the later stages of  a merger involving at least one
gas-rich galaxy \citep{Heckman86,Surace98}.

For the imaging part of the project we have used HST archive data
taken with various cameras and filters sensitive from the UV to the
near-IR. The wide spectral coverage of the observations allows us to
make accurate estimates of the ages of the stellar population
associated with the bright knots identified in the images. For the
spectroscopy  we have used high-quality long-slit spectra presented in
\cite{Holt03}. The spectroscopic data enable us to investigate the
ages and mass contributions of the YSP detected at optical
wavelengths in the diffuse halo of the
galaxy. We compare the results with model predictions
in order to understand the past and
future of PKS1345+12.

\section{OBSERVATIONS AND DATA REDUCTION}
\subsection{HST DATA}
The HST dataset comprises four sets of images taken with the FOC,
WFPC2, ACS and NICMOS cameras. A summary of the observations is
presented in Table 1.
 
\begin{table*}
\centering
\begin{minipage}{140mm}
\begin{tabular}{@{}llllllllcc@{}}
\hline 
Camera & &Filter & &$\lambda$$_c$($\AA$)(rest frame) & &$\Delta\lambda$($\AA$)(rest frame) & &Exposure Time(sec)\\ 
\hline 
FOC&  & F320W-POL0& & 2766& & 752& & 602 \\ 
FOC&  & F320W-POL60& & 2773& & 752& & 529 \\ 
FOC&  & F320W-POL120& & 2766& & 752& & 602 \\ 
WFPC2& & F439W& & 3844& & 423 & & 2200& \\ 
WFPC2& & F814W& & 7127& & 1567& & 835& \\ 
ACS(HRC)& & FR459M&  & 4539& &410 & & 2480& \\ 
ACS(HRC)& & F550M& & 4973& & 488& & 2480& \\ 
ACS(WFC)& & FR647M&  & 5897& & 530& & 800& \\
ACS(WFC)& & FR647M&  & 6549& & 588& & 800& \\ 
NICMOS& & F110W& &10060& & 5349& & 95& \\ 
NICMOS& & F160W& & 14316& & 3566& & 95& \\
\hline \hline
\end{tabular}
\caption{Summary of the HST observations of PKS1345+12 used in this
paper}
\end{minipage}
\end{table*}

\begin{table*}
\centering
\begin{minipage}{140mm}
\begin{tabular}{@{}llllllllll@{}}
\hline knot   &    &  \multicolumn{5}{r}{flux(/10$^{-18}$ erg
sec$^{-1}$ cm$^{-2}$ $\AA^{-1}$) }\\  
number    &F320W   &F439W  &FR459M &F550M$^{b}$   &FR647M  &F647M$^{b}$   &F814W  &F110W &F160W \\ 
\hline   
C1   &2.86$\pm$0.57   &2.09$\pm$0.15 &1.55$\pm$0.06   &1.35$\pm$0.05   &0.75$\pm$0.05   &1.30$\pm$0.05 &0.70$\pm$0.05 &$<$0.40   &$<$0.24 \\   
C2 &1.44$\pm$0.29 &1.68$\pm$0.12  &1.35$\pm$0.05   &1.11$\pm$0.04 &0.98$\pm$0.05 &1.43$\pm$0.06
&0.69$\pm$0.05   &$<$0.40 &$<$0.24 \\   
C3$^{a}$   & ...   & ...   &... & ...   &0.88$\pm$0.05   &1.35$\pm$0.05   &0.45$\pm$0.04   & ...
& ...\\   
C4   &1.56$\pm$0.30   &1.17$\pm$0.10   &0.71$\pm$0.03 &0.76$\pm$0.03   &0.58$\pm$0.04   &0.96$\pm$0.04   &0.35$\pm$0.04 &$<$0.40   &$<$0.20 \\  
\hline
\end{tabular}
\caption {Fluxes measured for the clusters at various wavelengths from
the HST images. \newline $^{a}$For C3 we were only able to obtain
accurate photometry for three photmetric bands, due to the proximity
of this cluster to one of the nuclei (see Fig 1). \newline $^{b}$These
two filters contain the [O\,{\sc iii}] and
H$\alpha$ emission lines. }
\end{minipage}
\end{table*}

The observations and data reduction of the FOC images are described in
\cite{Hurt99}. These pre-costar images were taken with the f/96 relay,
with a combination of the F320W filter and three different polarizing
filters (POL0, POL60 and POL120). The central wavelength of this
configuration is ${\lambda_c}=3100$~\AA\, for a power-law spectrum
$f_{\lambda}{\propto}{\lambda}^{-1}$.

For the purpose of this paper we are not interested in the
polarization, but  rather the total flux at various
wavelengths. Therefore we adapted the data to make them suitable for
this study. Since the three images were taken at the same wavelength,
we averaged them using the IRAF routine IMCOMBINE. Assuming that the
light from the bright knots is unpolarized ---a reasonable
assumption--- the polarizing filters will transmit half of the total
flux. Therefore the fluxes derived from these data using the standard
photometric calibration factors were  multiplied by a factor of two in
order to work out the total fluxes. The absolute photometric accuracy
of the f/96 relay is 10 -- 20~\% \cite{Baum94}.

\begin{figure*}
\vspace{3.2 in}
\begin{center}
\includegraphics{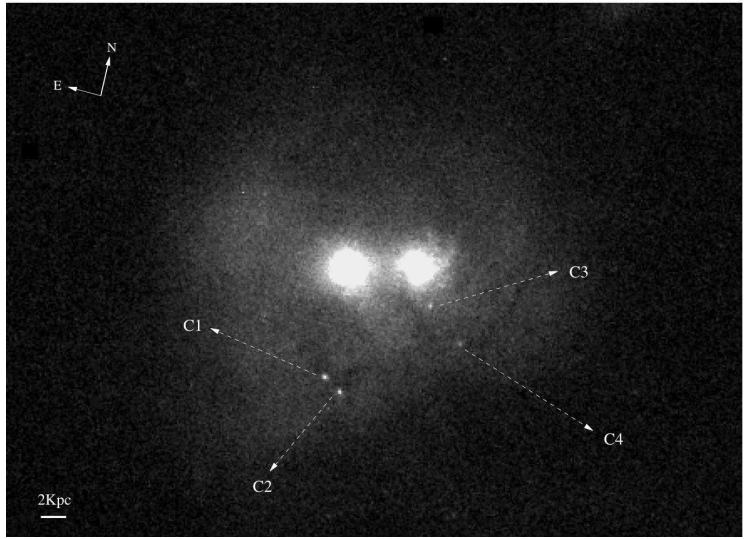}
\end{center}
\vspace{3mm}
\caption{The ACS $H\alpha$ continuum image of PKS1345+12 showing the
presence of two nuclei, a distorted halo and four bright compact
clusters embeded in this halo. The image also shows the 
presence of dust features spread to the
south of the galaxy. These dust features intercept
the positions of three of the clusters (C1,C2,C4).}
\end{figure*}

The WFPC2 observations and their reduction are described in detail by
\cite{Surace98}. The PKS1345+12 images were centred in the PC chip
($0.046$ arcseconds pixel$^{-1}$). The filters used were F439W and
F814W, which are very similar to the standard Johnson B and Cousins I
filters \citep{Holtzman95a, Holtzman95b}. Dark subtraction, bias
subtraction and flat fielding were carried out using the standard data
reduction pipeline procedures at STScI. The images were combined using
the IRAF/STSDAS routine CRREJ which removes cosmic rays in the
combined image. The calculated  errors on the photometric points,
including both the photon noise and calibration uncertainity, are
$\pm$7-10 per cent.

Four ACS images of PKS1345+12 are used in this paper. The Wide Field
Channel (WFC: $0.049$ arcseonds pixel$^{-1}$) was used in  combination
with FR647M medium band ramp filter to take two of the images, one
centred on the redshifted H$\alpha$ emission line wavelength
(hereafter H$\alpha$ image), and the other centred on the nearby
continuum (hereafter H$\alpha$ continuum image). The High Resolution
Channel (HRC: $0.027$ arcseconds pixel$^{-1}$) was used in combination
with  F550M medium band filter ($\lambda_{0}=5579$~\AA\,
$\Delta\lambda_{FWHM}=547$~\AA\,) and FR459M medium band ramp filter
to take the other images. In the former, the aim was to take an image
of the galaxy at the wavelength of the [O\,{\sc iii}]$\lambda5007$
emission line (hereafter O\,{\sc iii} image), whereas in the other
images the pivot wavelength was chosen to take an image of the
continuum near the [O\,{\sc iii}]$\lambda5007$ line ( hereafter
O\,{\sc iii} continuum image).  The FR647M($\lambda_c=6615$~\AA) image
is shown in Figure 1. The data were reduced following the standard
data reduction pipeline procedures at STScI which employ two packages:
CALACS package, which performs dark substaction, bias substraction and
flat field corrections producing calibrated images, and MULTIDRIZZLE
package which corrects for distortion and performs cosmic rays
rejections.  Since two of the ACS images were set up to measure the
H$\alpha$ and [O\,{\sc iii}]$\lambda5007$ emission lines,  they were
not suitable to be used for modelling the continuum, and only the
FR647M ($\lambda_{0}=6615$~\AA\,) and FR459M images were used for that
purpose. The calculated errors of the photometric points including
both the photon noise and calibration uncertanty are $\pm$5-7 per
cent.

For the FOC, WFPC2 and ACS data, the routine PHOT in IRAF was used
measure  sky-subtracted fluxes for the clusters within circular
apertures typically 3-10 pixels in radius, correcting for aperture
losses using a synthetic PSFs generated with the TINYTIM program. The
measured fluxes are summarised in Table 2.

The observations and reduction of the NICMOS data are described in
\cite{Scoville00}. The images were taken using camera 2 of NICMOS,
having a spatial resolution of $0.0762$ and $0.0755$ arcseconds
pixel$^{-1}$ in x and y respectively \citep{Thompson98}. The
F110W(1.10~$\mu$m, $\Delta\lambda_{FWHM} \sim 0.6$~$\mu$m),
F160W(1.60$\mu$m, $\Delta\lambda_{FWHM} \sim 0.4$~$\mu$m) filters were
used for the observations. The images we used for this paper were
fully reduced and calibrated using the standard pipeline procedure.

No SSCs are clearly visible in the NICMOS images. However, the images
are  useful for measuring upper limits, which provide 
constraints on the long wavelength SEDs. For that purpose, the TINYTIM
program was  used to create synthetic point source images for each
filter which were  scaled using a range of flux scaling factors. The
synthetic point sources were then added to the images at the expected
locations of the SSC, and the flux scale factor increased until the
point sources were just detected in visual inspection of the
images. The fluxes of the just-detected point sources were taken as
upper limits on the true near-IR fluxes of the SSC. We summarize the
results based on the NICMOS data in Table 2.

\subsection{Spectra}

In addition, we have used long-slit spectra to study the YSP
associated with  the diffuse light component in PKS1345+12. The
observations and reduction processes are described in
\cite{Holt03}. The spectra were taken in 2001 May with the ISIS
spectrograph on the 4.2m William Herschel Telescope (WHT) on La
Palma. The final wavelength range is 3275-6813~\AA\, in the blue and
6212-7720~\AA\, in the red. Useful spectra were taken along two slit
positions: PA 160 and PA 230 with a 1.3 arcsecond slit. The final
wavelength calibration accuracy was 0.059~\AA\, and 0.112~\AA\, in the
red and blue respectively, the spectral resolution is
3.66$\pm$0.09~\AA\, in the red and 4.54$\pm$0.10$\AA$ in the blue, and
the relative flux calibration accuracy is $\pm$5 per cent. A
preliminary analysis of this dataset for one aperture is presented in
\cite{Tadhunter05}, whereas for the current study several apertures
were extracted and analysed from both slit positions using the
STARLINK packages FIGARO and DIPSO. Prior to modelling, the spectra
were corrected for galactic reddening ($E(B-V) = 0.034$:
\cite{Schlegel98})  and we also subtracted the nebular continuum for
those apertures with strong emission lines
\citep{Dickson95,Holt03}. In the case of the nuclear aperture, where
we have a good estimate of the reddening of the various kinematic
components, the optimum reddening model of \cite{Holt03} was used to
generate a nebular continuum. However, for apertures 1 and 2 the
reddening is not known accurately because of the effect of the
underlying absorption lines on the high order Balmer lines. For those
apertures we used H$\alpha$ to generate the nebular continuum and
considered two extreme cases: (i) maximum nebular continuum assuming
no reddening of the emission line region; (ii) zero nebular continuum
corresponding to high reddening. In the remaining apertures (PA 230, A
\& B; PA 160, 3 \& 4) no nebular  continuum subtraction was deemed
necessary.  We then performed spectral synthesis modelling of all the
extracted spectra; the results are presented in next section.

\section{Results}
\subsection{Photometric Analysis}

The deep ACS H$\alpha$ continuum image presented in Figure 1 shows the
presence of the two nuclei of this double nucleus system, as well as a
common non-elliptical, distorted, halo of diffuse light surrounding
the galaxy. Four bright super star clusters (SSC)\footnote{Althought
these bright knots appear unresolved, at the distance of PKS1345+12 we
cannot rule out the idea that they represent multiple star cluster
systems rather than single SSC.} are clearly identified in the image and
are the subject of our photometric analysis. Dust lanes are also
detected as fine, extended features of lower surface brightness in the
extended halo to the south of the galaxy. Significantly, the most
prominent of these dust lanes intercept the positions of three of the
SSC (C1, C2, C4).  We further note that one of the clusters (C3) is
close to the western nucleus. Due to this proximity, the flux
measurements are less reliable for this cluster because of the large
gradient in the galaxy background light.  Moreover, for some of the
images this cluster is not clearly detected. Therefore, it has not
been considered in the modelling process.

Rather than the colour-colour or colour-magnitude diagrams that have
been used previously to determine the ages of SSC in ULIRGs from
multi-wavelength HST imaging observations \citep{Surace98,Wilson06},
we instead  adopt the approach of modelling the optical/UV  spectral
energy distributions (SEDs) derived from our photometric measurements.
The main advantage of this approach is that it utilises all of the
photometric information simultaneously, and thereby allows both the
reddening and age of each cluster to be determined, without the
age/reddening degeneracy that can be a problem for single
colour-colour and colour-magnitude diagrams.

Stellar population synthesis models in the GALAXEV (version 2003)
library \citep{Bruzual03} were used to estimate the ages and masses of
clusters C1, C2 and C4 from their respective SEDs. We used the  {\it
galaxevpl\/} program within the library to construct synthetic SSC
spectra with ages in the range 1Myr to 11Myr. These spectra were
created assuming an instantaneous burst of star formation, solar
metallicity and using the \cite{Salpeter55} IMF with lower and upper
mass cutoffs $m_L =0.1$~M$_{\odot}$ and
$m_U=100$~M$_{\odot}$. Reddened model spectra were created for the
reddening range $0.0 < E(B-V) < 1.0$  using the parametrized Galactic
extinction law of \cite{Seaton79}. The five photometric UV/optical
points suitable for the fit, i.e. those not affected by  emission line
contamination, were modelled using a minimum  $\chi^{2}$ technique to
find the ages of the SSC. Table 3 presents ranges of ages and
reddenings for which acceptable fits were obtained ($\chi^2_{red}
\lsim 1$); the ranges of measured masses are also presented. During
the fitting process, the same ($\pm$10$\%$) error was assumed for all
five photometric points. Taking in account the calibration error and
the photon noise, both already discussed  in section 2.1, plus small
shifts in the central wavelengths of the filters, and uncertainties
due to the gradient in the background near the clusters, we believe
that a mean error of $\pm$10$\%$ is realistic.

The best fitting models for the three clusters are shown in Figure
2. For comparison, some  models that fall outside of the range of
acceptable fits are also shown. The latter represent the models with
age/reddening combinations that give the best reduced chi-squared just
outside the acceptable zone. Taking the example of cluster C2 (see
Figure 2b)  the fit obtained for the 4~Myr with E(B-V)=0.5 model is
noticably better than that obtained for either the 3~Myr with
E(B-V)=0.6 or 6~Myr with E(B-V)=0.4 models. Moreover $\chi^{2}_{red}$
values found for YSP ages of 4-5~Myr and reddening of E(B-V)=0.5 are
$\lsim 1$, while those found outside this range of ages are $\gsim
2$. Therefore,  the range of ages considered to be valid for the case
of C2 is 4-5~Myr,  with the best fit found for a template of 4~Myr
with a reddening of E(B-V)=0.5. Note that the high fluxes observed for
the FOC F320W point for the three clusters, rule out starburst ages
above 50 Myr, even taking in account any reddening effect, since the
observed $U-B$ is already bluer than such a starburst.

Further evidence for the young ages measured for the SSC in PKS1345+12
is provided by the detection of emission lines from the clusters in
both the HST images (Figure 2) and the spectra (see next
section). Based on the fluxes determined from our H$\alpha$ emission
line images, and correcting for both the SSC continuum and a 40\%
contribution from [NII]$\lambda\lambda$6548,6584 emission lines in the
filter bandpass, we determine H$\alpha$ equivalent widths of
250$\pm$37, 370$\pm$34 and 606$\pm$86~\AA\, for clusters C1, C2 and C4
respectively. The instantaneous burst models of \cite{Leitherer99}
show that such high equivalent widths are only attained for clusters
with ages $t_{cl} < 6$~Myr, regardless of the assumed IMF. Since the
ISM in the clusters may not absorb all the ionizing photons generated
by the OB stars (as assumed by the Leitherer et al.  1999 models), the
ages could be significantly less than this upper limit. Thus the
detection of H$\alpha$ emission lines with high equivalent width is
consistent with the range of ages determined from  SED modelling of
the clusters\footnote{Note that, although  we have not bee able to
make detailed fits to the continuum SED of cluster C3, it is clear
from Table 1 that this cluster also has H$\alpha$ detected at high
equivalent width, and therefore it is also relatively young ($t_{cl} <
6$~Myr).}. Finally we note that the NICMOS upper limits  rule out ages
older than $\sim$11~Myr with reddenings similar to or lower than those
presented in Table 3, or ages $<$7~Myr with higher reddening values
than the models in Table 3.

The masses of the SSC detected in PKS1345+12 ($10^6 < M_{ssc} <
10^7$~M$_{\odot}$) are comparable with those of the most massive
clusters detected in the Milky Way  (e.g. Omega Centauri), in merger
systems such as the ULIRG Arp220 \citep{Wilson06} and The Antennae
\citep{Fall05},  and in some other interacting systems
\citep{deGrijs}. Given the difficulty of detecting individual SSC at
the  relatively high redshift of PKS1345+12, it is not surprising that
the clusters we have detected are relatively young and
massive. Certainly, our imaging observations do not rule out the
presence of a more numerous cluster population in the halo of
PKS1345+12, comprising SSC that  are older and/or less massive  than
C1, C2 and C4.

\begin{table}
\centering
\caption{The results of the spectral synthesis
modelling applied to the optical/UV photometric points of Table 2.}
\begin{tabular}{ccccr}
\hline 
knot   &Age(Myr)    &E(B-V)   &Mass(10$^{6}$$M_{\odot}$) \\ 
number & & & &\\ 
\hline 
C1   &$<$7   &0.2-0.5      &$\sim$2.5 - 9.0     \\ 
C2   &4-5   &0.5          &$\sim$4.0 - 4.8     \\ 
C4   &$<$7   &0.2-0.5      &$\sim$1.3 - 5.0     \\ 
\hline
\end{tabular}
\end{table}

\begin{figure*}
\begin{center}
\psfig{figure=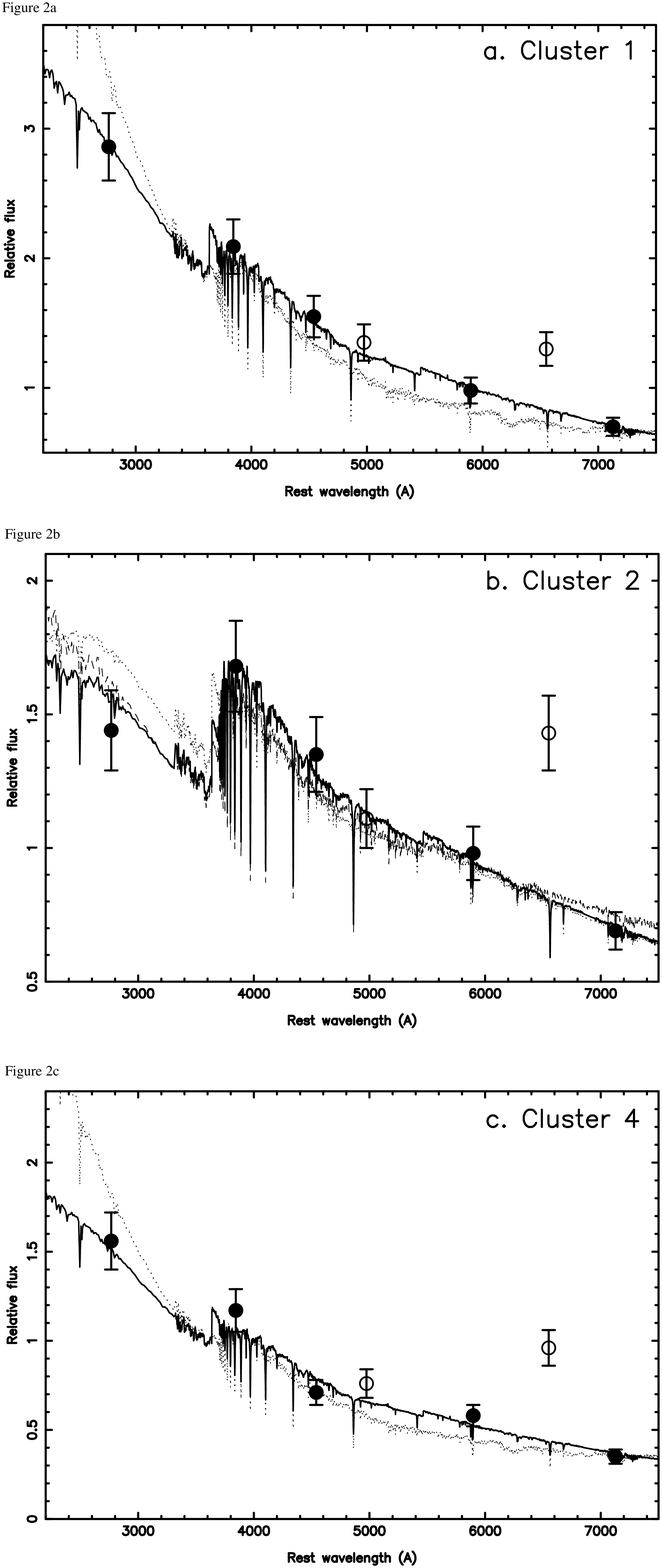,width=9.cm,angle=0.}
\end{center}
\caption{The results of fitting different templates to the fluxes
  measured for the three clusters used in the photometric analysis
  (flux measurements shown by solid symbols). Figure 2a shows the fits
  obtained for C1  using templates of 2 Myr with $E(B-V)=0.5$ (solid
  line) and 8~Myr with $E(B-V)=0.1$ (dotted line). It is clear in the
  figure that a better fit is obtained for the template of
  2~Myr. Figure 2b shows the plots obtained after fitting a 3~Myr
  template with $E(B-V)=0.6$ (dotted line), 4~Myr with $E(B-V)=0.5$
  (solid line), and 6~Myr with $E(B-V)=0.4$ (dashed line) to the data
  of C2. In this case it is clear that the best fit is obtained for
  the 4~Myr template. Figure 2c shows the results of fitting templates
  of 2 Myr with $E(B-V)=0.5$ (solid line) and 8~Myr with $E(B-V)=0.1$
  (dotted line) to the C4 data. The best fit this time is found for
  the 2~Myr template. The flux measurements obtained for H$\alpha$ and
  [O\,{\sc iii}] emission lines are also shown in the figure with open
  symbols. The fluxes are presented in wavelength units}
\end{figure*}

\begin{figure*}
\begin{center}
\psfig{figure=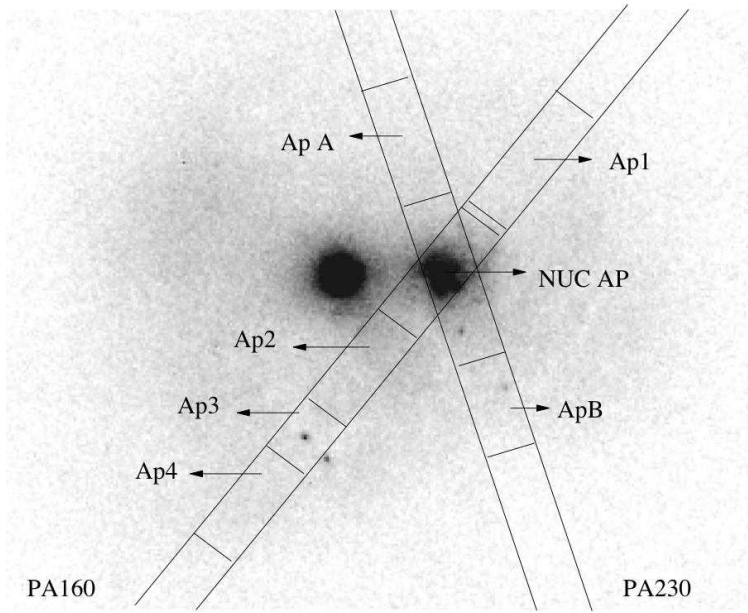,width=10.cm,angle=0.}
\end{center}
\caption{H$\alpha$ continuum image showing the locations of the slits and
the extraction apertures.}
\end{figure*}
\begin{figure*}
\begin{center}
\psfig{figure=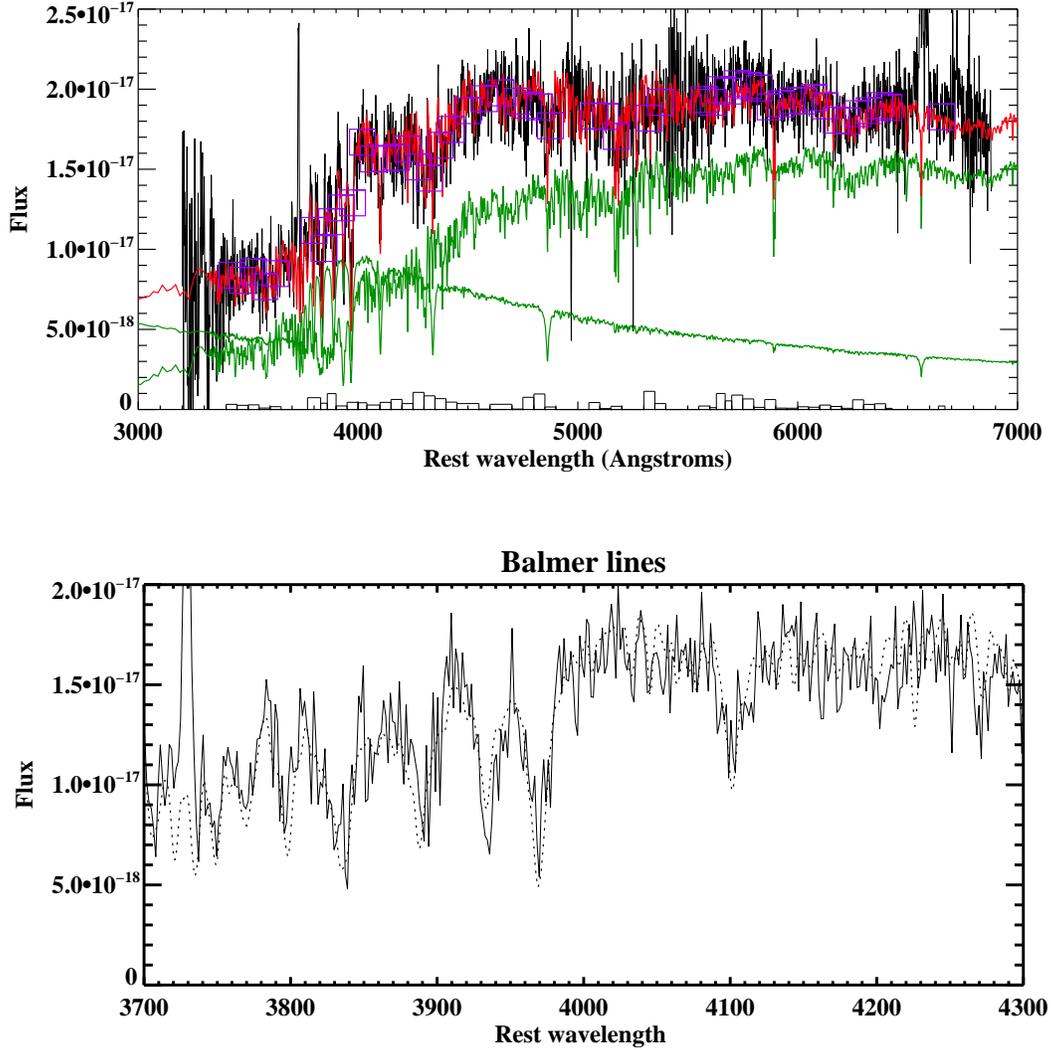,width=14.cm,angle=0.}
\end{center}
\caption{Upper panel: modelling result for the spectum extracted from
  Ap4 of Fig 3. The green lines represent the young and old templates,
  while the red line represent their sum. The purple boxes represent
  the bins used for the fit, and the histogram on the x-axis
  represents the modulus of the error measured in each bin. In this
  case the young component has an age of 0.2 Gyr with a reddening of
  $E(B-V)=0.1$. The contribution of this component to the total
  flux(red line) is 34.6 $\%$ in the normalising bin (4600-4700\AA)and
  the $\chi^{2}_{red}$ in this case is 0.4. Lower panel: detailed fit
  in the wavelength range 3700-4300. The fluxes are presented in
  wavelength units}
\end{figure*}
\begin{figure*}
\begin{center}
\psfig{figure=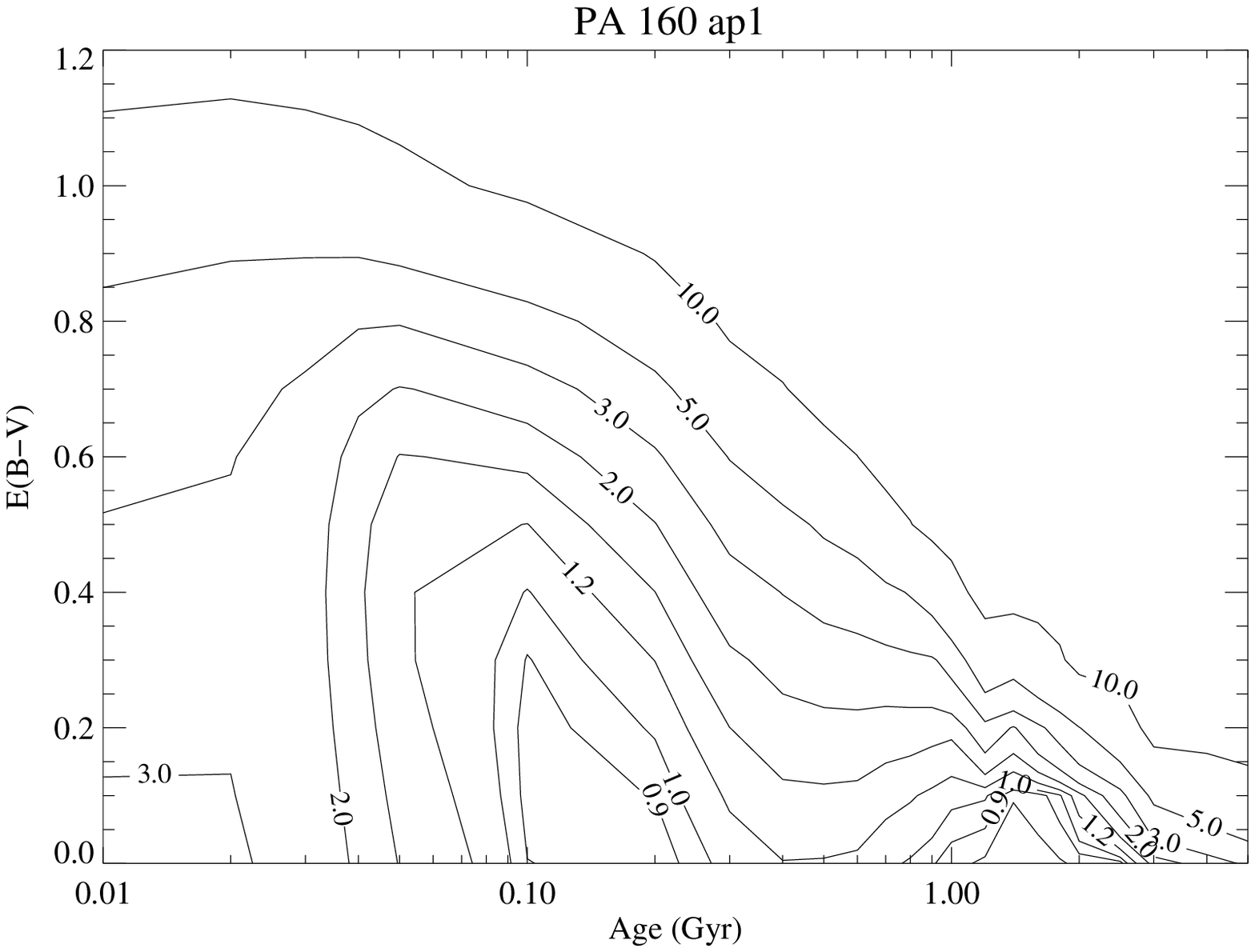,width=10.cm,angle=0.}
\end{center}
\caption{Minimum chi-squared fitting results for Ap1 in Figure 3. Each
contour represents a value for $\chi^{2}_{red}$. Good results for the
modelling are considered $\chi^{2}_{red}$ of one or less (see
discussion in Tadhunter et al. 2005). The two loci where good results
are found are placed around 0.1 Gyr with $E(B-V)= 0.2$ and 1.5 Gyr with no
reddening}
\end{figure*}

\subsection{Analysis of the Spectra}

Figure 3 shows the slit positions and the apertures used for the
spectroscopic analysis. The same routine as described above was used
to create simple stellar population templates of different age and
reddening. Reddened model spectra were created using E(B-V) values
from 0.0 to 1.6 and the parametrized galactic extinction law of
\cite{Seaton79}. Each extracted spectrum was modelled using a
combination of a young and an old stellar population (see
\cite{Tadhunter05}): we used a 12.5Gyr model for the old stellar
population (OSP), adding YSP ranging in age from 0.001Gyr to 6Gyr (we will
refer to this component as young component). For each spectroscopic
aperture the continuum flux was measured in several wavelength bins
($\sim$50) chosen to be as evenly distributed in wavelength as
possible, and to avoid strong emission lines and atmospheric
absorption features.  We then fitted these data using a minimum
$\chi^2$ technique to determine the best fitting age and reddening, as
well as the relative contributions of each of the two populations. The
result  of the modelling the spectrum extracted for aperture Ap4 is
shown in Figure 4 as an example.

Although the spectra have the advantage of considerably better
wavelength sampling than the HST images, they have the disadvantage
that several stellar populations of different age/reddening may be
included in the large spectroscopic aperture. This can lead to an
age-reddening degeneracy in the sense that adequate fits can sometimes
be obtained for either a small proportion of a relatively young,
highly reddened YSP plus an old stellar population,  or a much larger
proportion of an intermediate age, but less reddened YSP.  Figure 5
presents the details of the mininmum $\chi_{red}^{2}$ modelling for
aperture Ap1 and provides a good illustration this degeneracy problem
(see also Tadhunter et al. 2005), since the two loci that represent
adequate fits ($\chi_{red}^{2}\lsim 1$) are clearly separated in the
figure: at around $\sim$0.1~Gyr with E(B-V) = 0.2 and  $\sim$1.5 Gyr
with E(B-V) = 0.0. To distinguish between groups of templates that
provide good fits, the model fits to the data were examined in detail,
selecting those models which provide the best fit to the important
absortion features of the spectra (for example CaII~K, G-band, higher
order Balmer lines).  Figure 6 shows the wavelength range  3700\AA\
--- 4300\AA\ for two of the best solutions obtained from the SED
modelling of Ap3, PA160. It is clear from the plot, just by examining
the features of the fit for the CaII~H\&K absortion lines, that in
this case the model comprising 12.5~Gyr OSP plus the 0.04~Gyr YSP
component age fits the data better than the 1.4 Gyr YSP
model. However, there are cases (Ap2, Ap3, ApB) where even younger
($\sim$0.005~Gyr) age YSP --- corresponding to the
blue A/B supergiant phase ---  provide an adequate fit to the
SED. For these latter cases we cannot distinguish between YSP with
ages $\sim$0.005~Gyr and $\sim$0.04~Gyr using the detailed fits.

Table 4 sumarizes the results of the modelling for all of the
extracted spectroscopic apertures. We find ages for the YSP ranging
from 0.005 Gyr to 1.5 Gyr with reddenings in the range of E(B-V)=0.0
to 1.2. We also measured the mass contribution of the YSPs to the
overall stellar mass in the aperture, finding results ranging from a
negligible mass contribution ($< 1$\%) in some apertures, to the YSP
contributing a sizeable fraction of the total mass in others. The wide range
of ages found for some of the apertures reflects the fact that good
fits can also be achieved for two combinations of age/reddening (the
degeneracy problem discussed above), and also depends on the nebular
continuum subtraction (for Ap1 and Ap2). The reader may notice that
the results for NUC AP in Figure 3 are not presented in Table 4. After
the nebular subtraction, the flux contribution of the YSP in the
normalising bin is small ($< 10$\%). Therefore we find good fits for
the entire range of combinations OSP and YSP templates.

It is notable that for the region to the south of
the nucleus that encompasses the SSC detected in the HST images
(i.e. Ap2, Ap3 and ApB ), the minimum age that provides an
adequate fit to the SED of the diffuse light ($\sim$0.005~Gyr) is
consistent with the ages of the SSC measured in our photometric study,
although older YSP ages ($>$0.04~Gyr) are also acceptable. On the
other hand, we can {\it only} obtain adequate fits to the SEDs of the
other extended apertures (Ap1, ApA \& Ap4) using older YSP with ages
in the range 0.04 -- 1.0~Gyr, and we cannot rule out the idea that YSP
sampled by the diffuse light have the same age ($\sim$0.1 -- 0.3~Gyr)
for all apertures.

Overall, the range of YSP ages we  have
determined for the diffuse light from our spectra of PKS1345+12
overlaps with the 1 -- 300~Myr range determined for the YSP in the
spectroscopic study of a sample of transitional ULIRG QSOs by Canalizo
et al.  (2001); it is also consistent with the 0.5 -- 1.5~Gyr age
determined by Tadhunter et al. (2005) for a single extended aperture
to the SE of the nucleus of PKS1345+12 itself, although this latter
study considered only unreddened YSP.

Finally we emphasise that the diffuse light contains the dominant YSP
component in terms of mass. A lower limit on the total mass in the
diffuse  YSP component integrated across all our spectroscopic
apertures is  $M^{YSP}_{diff} > 1.4\times10^9$~$M_{\odot}$, and may be
considerably larger if the regions of the galaxy not sampled by the
slits are included. In contrast,  the  total mass in the three SSC
with accurate age estimates is orders of magnitude less
($8\times10^6 < M ^{YSP}_{ssc} < 2\times10^7$~M$_{\odot}$).

\begin{table*}
\centering
\begin{tabular}{llllllllllllll}
\hline
&  &Age of   &   &E(B-V)$^a$   &   &$\%$YSP\\
&  & YSP     &   &         &    &of total\\
&  &(Gyr)    &   &         &    &mass\\
\hline
PA 160    &Ap1   &0.1 - 0.3   &   &0.0 - 0.4          &  &1.0 - 2.0\\
          &Ap1-neb$^b$   &0.2 - 1.2   &   &0.0 - 0.4  &  &1 - 69\\
          &Ap2   &0.001 - 0.2      &   &0.0 - 1.2      &  &0.05 - 3.5\\
          &Ap2-neb$^b$   &0.2 - 1.0   &   &0.3 - 0.7  &  &1 - 60 \\
          &Ap3   &0.005 or 0.04 - 0.2  &   &0.4 -0.8 or 0.0 - 0.5          &  &0.1 - 1.5 or 1.5 - 15.0\\
          &Ap4   &0.05 - 0.5   &   &0.0 - 0.6         &  &2 - 10\\
\hline
PA 230    &ApA   &0.05 -1.5    &   &0.0 - 0.4         &  &2.0 - 90.0\\
          &ApB   &0.005 or 0.04 - 1.5  &   &0.0 - 1.0 or 0.0 - 0.7          &  &0.01- 1.2 or 7.0 - 90.0\\
\hline
\end{tabular}
\caption{Results from the analysis of the spectra extracted from the
apertures in Figue 3.\newline  $^a$ The large range in reddening
should be understood as it follows. For the younger solutions there is
a wide  range of reddenings which gives good results. However, when
moving towards older ages the amount of reddening needed for the
template to match the data decreases as does the range. \newline $^b$
Two different sets of results are presented for the apertures 1 and
2. One is obtained from modelling without subtracting any nebular
continuum from the spectra, while the other (labelled Ap1-neb and
Ap2-neb) is obtained after having subtracted the maximum nebular
continuum. Thus the true results are likely to fall between the the
resuls presented in the table}
\end{table*}
\begin{figure*}
\begin{center}
\psfig{figure=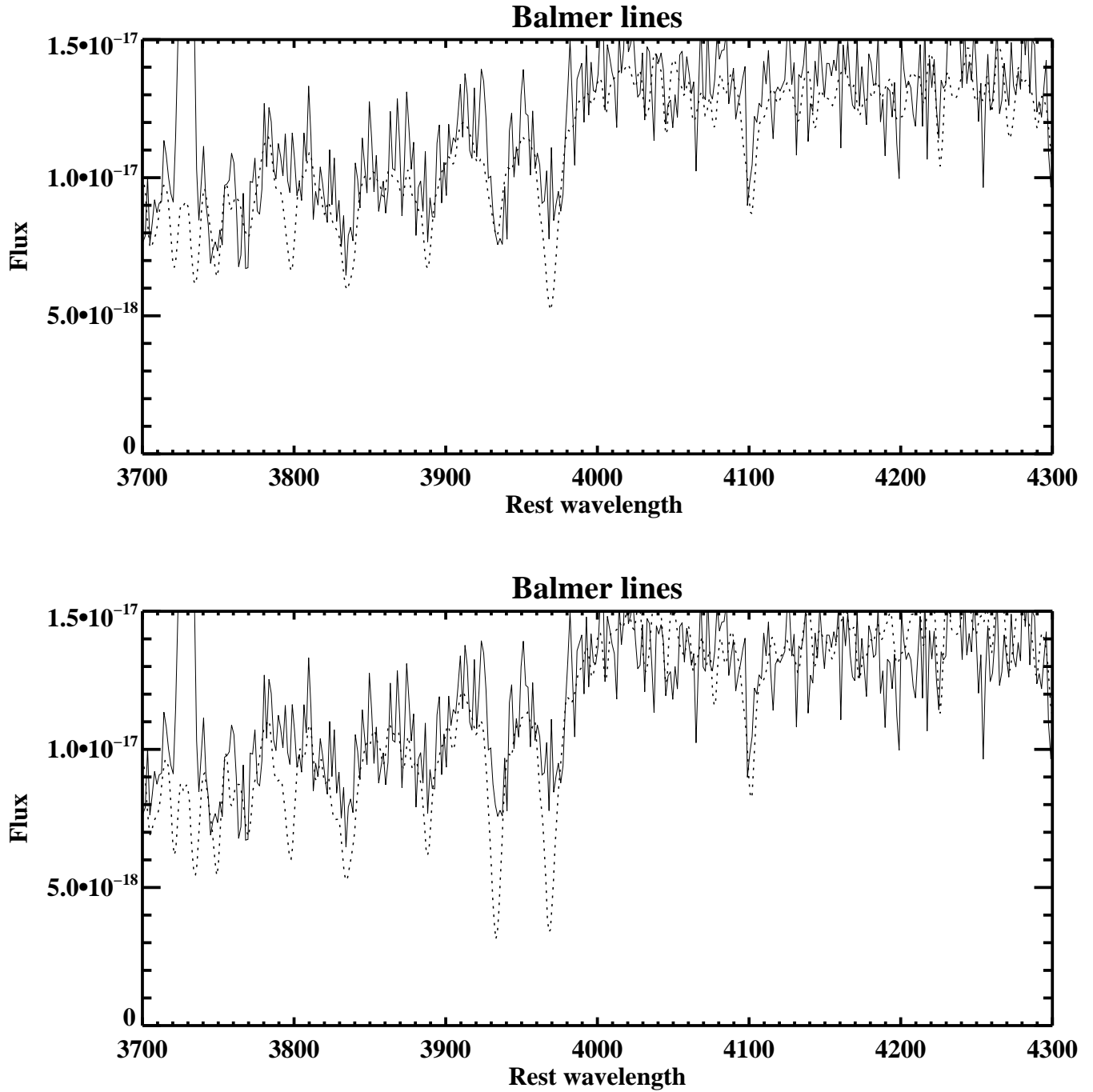,width=18.cm,angle=0.}
\end{center}
\caption{Detailed fits for the spectrum extracted from Ap3. Upper
panel: detailed fit for a model with a young stellar component of 0.04
Gyr (E(B-V)=0.2). Lower panel: detailed fit for a model with a
``young'' stellar component of 1.4 Gyr (no reddening). The fluxes are
presented in wavelength units. It is clear from the figure
that the model with the younger YSP (upper panel) fits the absortion
features best, despite the fact that both models provide adequate fits
to the overall SEDs.}
\end{figure*}

\subsection{Gas Kinematics in the Halo}

\begin{figure*}
\begin{center}
\psfig{figure=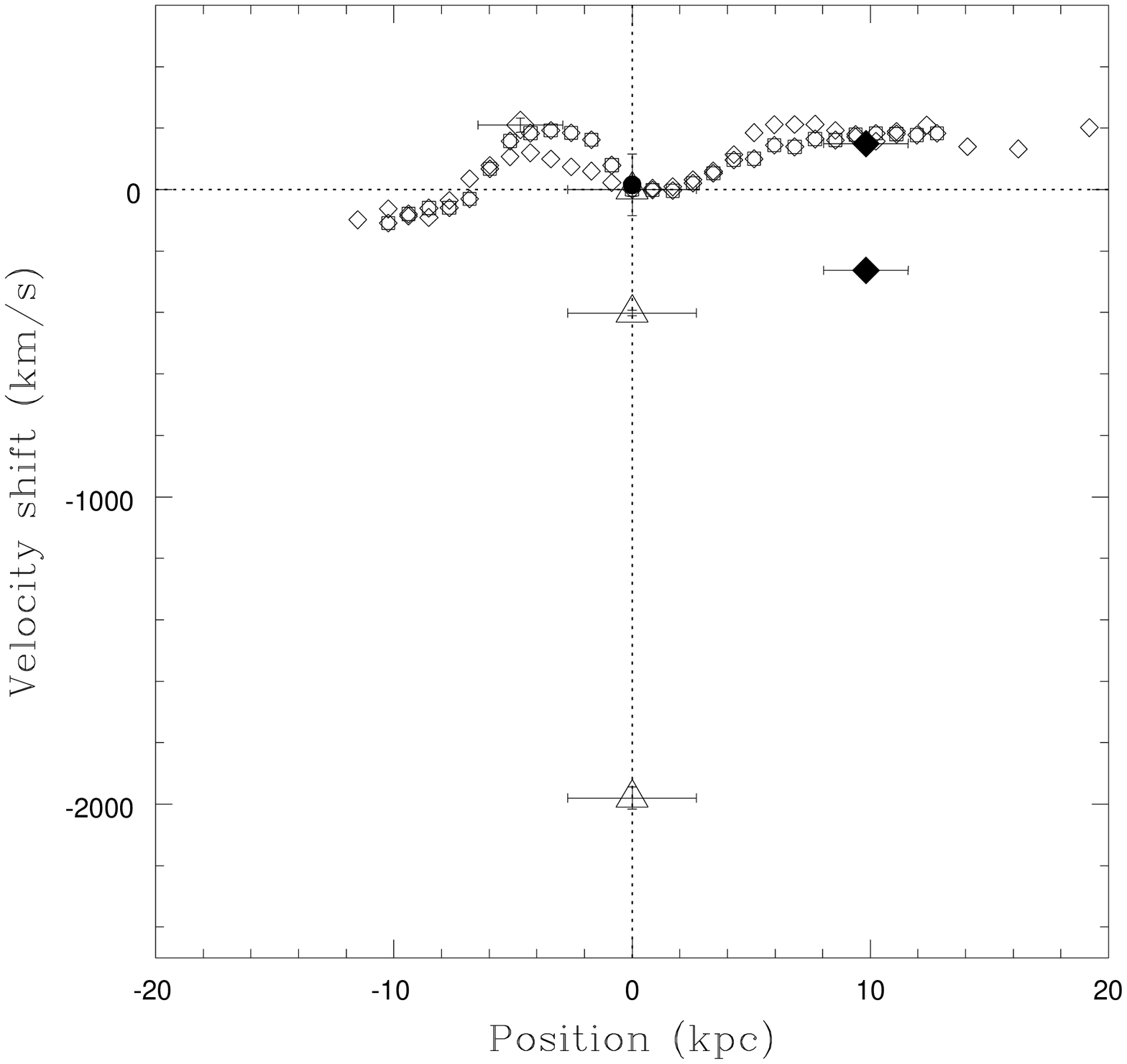,width=12.cm,angle=0.}
\end{center}
\caption{Spatial variations of the [O\,{\sc ii}]  and H$\alpha$ radial
velocities of the extended gaseous halo along PA 160. [O\,{\sc
ii}](open diamonds), H$\alpha$ (open diamonds + open squares). The
error bars are small and are incorporated in the points. Overplotted
is the radial velocity of the H\,{\sc i}~21cm absorption (filled
circle; \citet{Mirabel89}), the three components of [O\,{\sc iii}] in
the nucleus (open triangles) and the extended emission line regions:
the extended [O\,{\sc iii}] emission in the region to the NW (large
open diamond) and the extended H$\alpha$ emission in the line
splitting region to the SE coinciding with the position of the cluster
C1 (large filled diamonds). Note: Positive kpc = south, negative kpc =
north. The vertical dotted line represents the centroid of our nuclear
aperture. The horizontal dotted line represents the velocity at the
position of the centroid.}
\end{figure*}

The emission line kinematics of PKS1345+12 have been described in
detail by Holt et al. (2003). As well as a quiescent emission line
halo with relatively small radial velocity amplitudes ($<$400 km
s$^{-1}$) and linewidths ($<$500 km s$^{-1}$ FWHM). \cite{Holt03}
found extreme kinematic components at the location of the western
nucleus with velocity shifts up to $\sim$2000 km s$^{-1}$  and line
widths of  $\sim$2000 km s$^{-1}$ (FWHM) providing evidence for warm
gas outflows in the narrow line region. As mentioned in Section 3.1,
we detect emission lines at the locations of the young star clusters
visible in our HST in both the images and the long slit spectra.
Therefore it is interesting to compare the kinematics of the star
clusters with those of the other kinematic components in the galaxy.

Figures 7 and 8 compare the emission line kinematics at the locations
of the clusters with those of the spatially resolved diffuse gas
detected in the H$\alpha$ and [OII]$\lambda$3727 emission lines in the
halo of the galaxy along PA 160 and PA 230. Intriguingly, we find
clear evidence for line splitting in both H$\alpha$ and [NII]  at the
locations of clusters C1/C2 and C4. The line splitting is clearly
visible in the extracted spectra of these regions shown in Figures 9
and 10 and also the grey scale representation of the long-slit spectra
shown in Figure 11.  Important features of the line splitting regions
include the following.

\begin{itemize}
\item {\bf Linewidths.} In both regions there is a relatively broad kinematic
component  ($320 < FWHM < 400$ km s$^{-1}$ ) along with  a
narrower  kinematic component ($FWHM < 150$ km s$^{-1}$) that is spectrally
unresolved.
\item {\bf Radial velocity shifts.} In both regions the broader
component has a radial velocity  and line width consistent with that
of the diffuse gas at  larger and smaller radial distances from the
nucleus, whereas the narrower component  is significantly shifted
relative to the diffuse gas: {\it blueshifted} by $\sim$450 km
s$^{-1}$ for region C1/C2, and {\it redshifted} by $\sim$300  km
s$^{-1}$ for region C4.
\item {\bf Ionization state.} Whereas the narrow kinematic component
has relatively small [NII]6584/H$\alpha$ and [OIII]/H$\beta$ ratios
([NII]6584/H$\alpha <$ 0.5 and [OIII]/H$\beta <$ 1), consistent with
photoionization by hot stars in an HII region (see Veilleux et
al. 1999), the broader component has line ratios ([NII]6584/H$\alpha
>$ 0.7 and [OIII]/H$\beta <$ 3) that are more consistent with LINER
nuclei and the diffuse gas at other locations in the host
galaxy. Overall the LINER-like line ratios for the broader diffuse gas
component indicate ionization by shocks, or photoionization by an AGN
at low ionization parameter.
\end{itemize}

All of these features are consistent with the idea that the narrow
emission line components represent HII regions associated with the
young star clusters, and that these SSC are moving at high velocity
relative to both the ambient gas and the rest frame of the galaxy
defined by the narrow emission lines and the HI 21cm absorption line
detected in the nucleus. In this case, we are observing the clusters
before the phase in which the supernovae associated with the massive
stars in the clusters  eject the ISM
\citep{Goodwin97a,Goodwin97b,Bastian06}. Indeed, given that the
supernova phase is expected to start $\sim$3 -- 4~Myr after the birth
of a star cluster (Leitherer et al. 1999), the ages we  have
determined for the SSC ($\lsim 6$~Myr) are consistent with such a
supposition.  Given their measured radial velocities and estimated
ages, the clusters must have formed within a few kpc of their current
locations, unless their tangential velocities are considerably larger
than their radial velocities. This latter feature rules out the idea
that the clusters formed at much larger radii in the outer halo of the
merging system, then fell into the nuclear regions; {\it the gas out
of which the SSC formed must have been moving rapidly prior to the
formation of the star clusters}.

The alternative explanation, that we have caught the SSC in the
ejection phase, and  that the line splitting is caused by the ejection
of gas by SSC formed in the quiescent, diffuse gas
\citep{Goodwin97a,Goodwin97b,Bastian06} is difficult to reconcile with
the observations. In particular, in such a case it is not  clear why
the component apparently at rest relative to the ambient gas has
relatively broad lines and a [NII]6548/H$\alpha$ ratio consistent with
shocks or AGN photoionization, whereas the component shifted relative
to the rest frame (presumably the ejected component in this scenario)
has narrow lines and  an HII region-like [NII]/H$\alpha$ ratio
characteristic of stellar photoionization; one would expect the
reverse to be the case.

Finally we note that, despite the apparently ``free-floating'' status
indicated by their emission line kinematics, the SSC/HII regions are
linked with a more extensive interstellar medium (ISM) in the host
galaxies. First, there is clear morphological evidence from our high
resolution ACS images that the young star clusters are associated with
dust lanes in the host galaxy (see Figure 1).  Second, at the location
of C1/C2, the greyscale representation of the long-slit spectrum (see
Figure 11) shows a clear enhancement in the flux of {\it both} the
broad and narrow components, suggesting  a link between the HII
regions and the more diffuse gas.

\begin{figure*}
\begin{center}
\psfig{figure=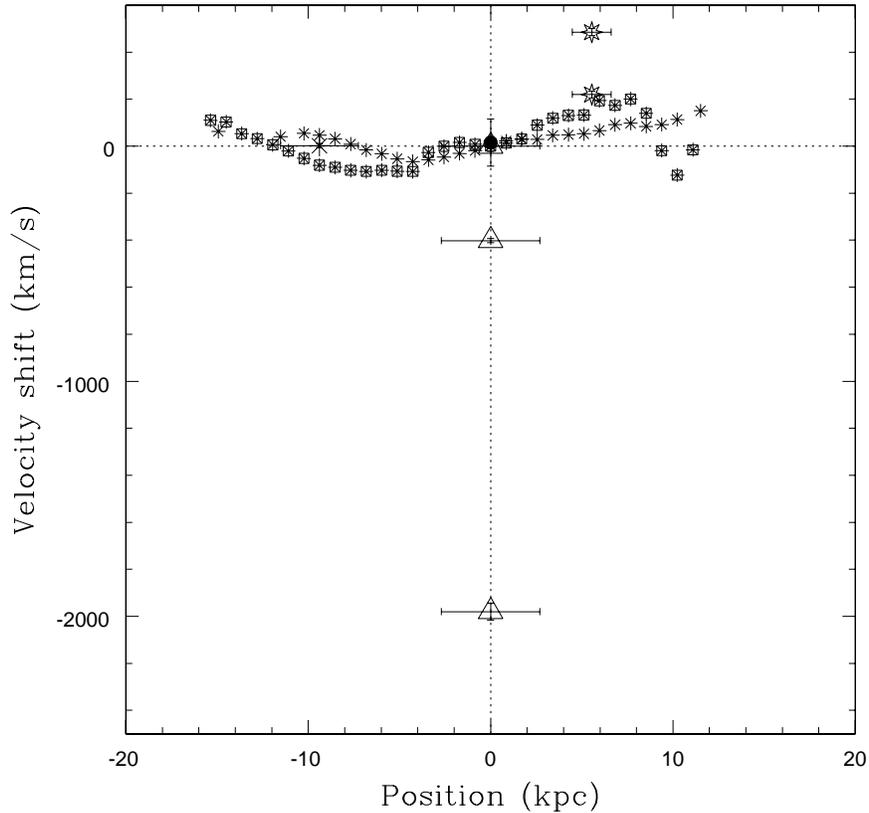,width=12.cm,angle=0.}
\end{center}
\caption{Same figure than 8 but for PA 230. [O\,{\sc ii}](asterisks),
H$\alpha$ (asterisks+ open squares). The error bars are small and are
incorporated in the points. Overplotted is the radial velocity of the
H\,{\sc i}~21cm absortion (filled circle; \citet{Mirabel89}), the three
components of [O\,{\sc iii}] in the nucleus (open triangles) and the
extended emission line regions: the extended [O\,{\sc iii}] emission
(large asterisk) and the extended H$\alpha$ emission in the line
splitting region to the SW coinciding with the position of the cluster
C4 (large stars). Note: Positive kpc = south-west, negative kpc =
north-east. The vertical dotted line represents the centroid of our
nuclear aperture. The horizontal dotted line represents the velocity
at the position of the centroid.}
\end{figure*}

\begin{figure*}
\begin{center}
\psfig{figure=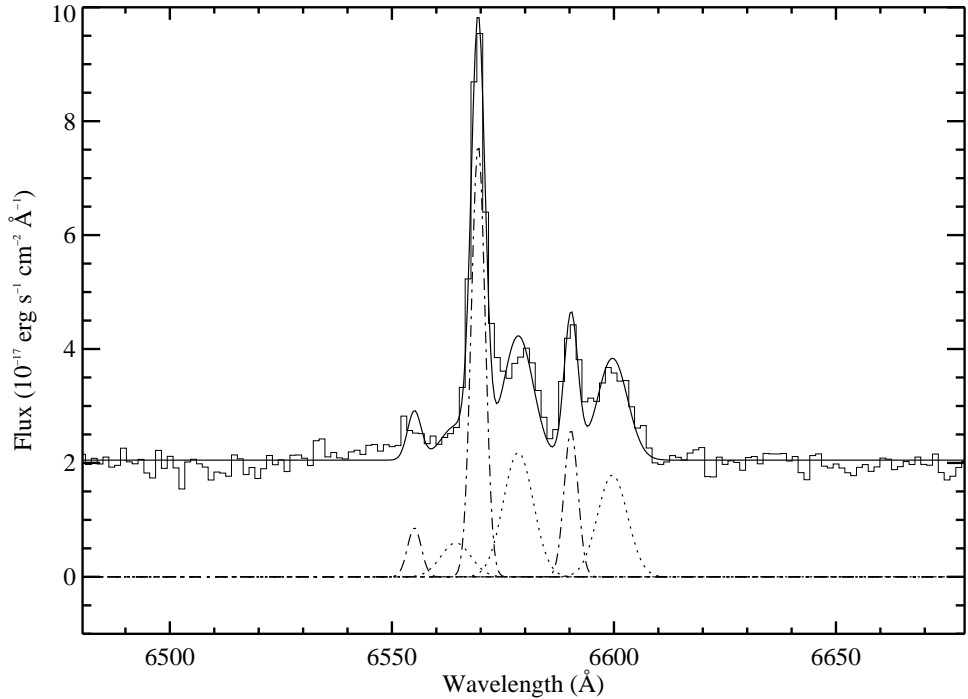,width=14.cm,angle=0.}
\end{center}
\caption{H$\alpha$ model of the region exhibitting line splitting
coinciding with the position of C1. The faint line traces the observed
spectra and the bold line represents the overall model profile. The
six components, two for each line, are also overplotted --- the dotted
line traces the broader component consistent with the kinematics of
the emission line gas at other locations in the galaxy, and the
dot-dashed line traces the  narrow component blueshifted by $\sim$450
km s$^{-1}$ with respect to the ambient gas.}
\end{figure*}

\begin{figure*}
\begin{center}
\psfig{figure=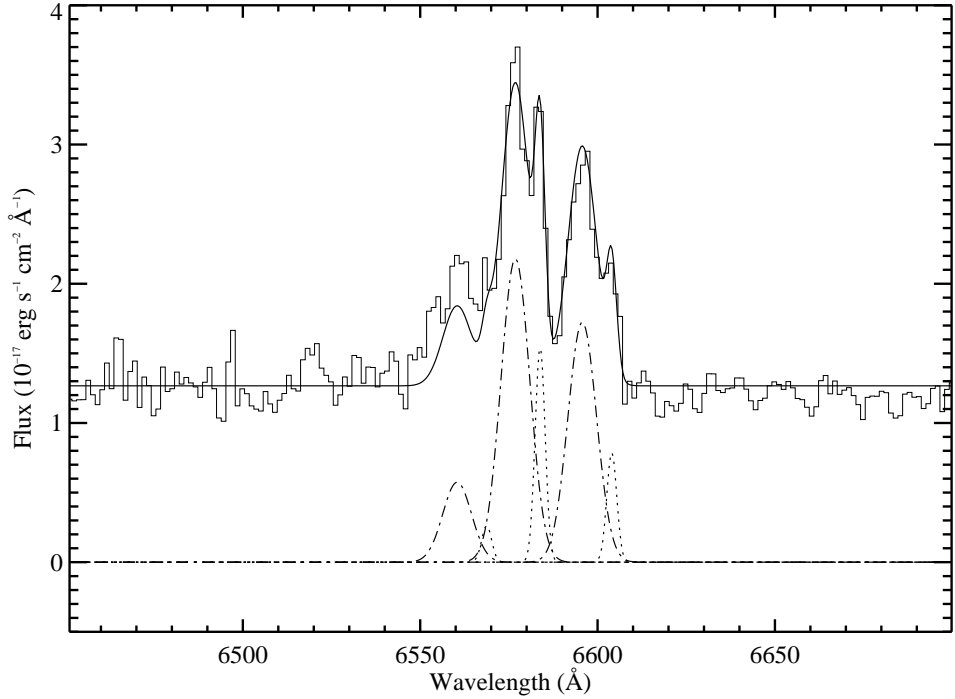,width=14.cm,angle=0.}
\end{center}
\caption{H$\alpha$ model of the region exhibitting line splitting
coinciding with the position of C4. The faint line traces the observed
spectra and the bold line represents the overall model profile. The
six components, two for each line, are also overplotted --- the dotted
line traces the broader component, and the dot-dashed line trace the
second narrow component redshifted by $\sim$300 km s$^{-1}$ with
respect to the spatially extended narrow component}
\end{figure*}

\begin{figure}
\vspace{1.50 in}
\begin{center}
\includegraphics{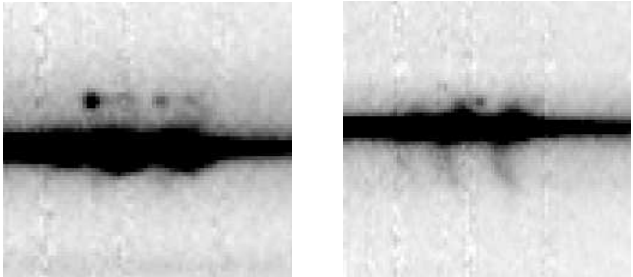}
\end{center}
\vspace{3mm}
\caption{2-D optical spectra for PA 160(left) and PA 230(right),
centred on the redshifted H$\alpha$+[NII] emission lines. The x axis
is the wavelength direction increasing to the right, and the vertical
axis is the spatial direction with SE to the top and NW to the
bottom. The images show the regions exhibiting H$\alpha$ line
spitting, coincident with the SSC, and demonstrate the flux
enhancement of both kinematic in the line splitting region in PA 160
(at the location of C1/C2).}
\end{figure}

\section{Discussion}
\subsection{Age determination for the YSPs in ULIRGs}

It is interesting to compare our results for PKS1345+12 with previous
studies of YSP in  ULIRGs.  \cite{Surace98} carried out an HST imaging
study concentrating on bright star forming knots for a sample of nine
``warm'' ULIRGs, including PKS1345+12, using high resolution B- and
I-band images taken with the Wide Field Planetary Camera on the HST.
The stellar synthesis models of \cite{Bruzual93} were used to estimate
the ages and masses of the bright knots in colour-magnitude diagrams.
However, It is difficult to distinguish between stellar age and
reddening using only B- and I-band observations. Thus the ages
presented in \cite{Surace98} are upper limits and, as discussed in that
paper, the effect of correcting for the reddening could
substantially modify the age and mass estimates, resulting in younger
ages and smaller masses. For the whole sample, they derive a median
upper limit for the age of $\sim3\times10^8$~yr and masses in the
range $10^5 < M_{clusters} < 10^9$~M$_{\odot}$. In the case of
PKS1345+12 itself, they found ages $\sim10^7 < t_{ssc} <  10^8$~yr and
masses $\sim10^7 < M_{clusters} < 10^8$ M$_{\odot}$. Since we have
measured significant amounts of reddening for the knots in PKS1345+12,
this galaxy is clearly a case in which the reddening affects the
results presented in \cite{Surace98}. As expected, after taking into
account the reddening, we deduce younger ages and lower  masses for
the SSC in PKS1345+12.

It is well known that large amounts of gas and dust surround the AGN
in the western nucleus of PKS1345+12
\citep{Evans99,Surace99,Scoville00,Holt03}. Our new results
demonstrate the importance of dust on a larger, $\sim$5 -- 15~kpc
scale, in the halo of  the galaxy. Not only do the SED fits require a
reddening  in the range $0.2 < E(B-V) < 0.5$, but there are dust
structures in our images coincident with the SSC. Therefore it  is
dangerous to use  single colour-magnitude diagrams to determine the
ages of the YSPs in such objects.

As well as the importance of taking into account  reddening, our
results also demonstrate that imaging and spectroscopic data provide
complementary information about the YSPs; clearly a combination of the
two types of data is required to make a full  census of the YSPs
present in systems of this type.  Whereas on the basis of photometric
analysis we find robust ages $<$6~Myr for the three SSC with good
SEDs, our spectroscopic analysis  places a lower limit of 40Myr on the
ages of the YSP associated with the diffuse light in three of
the regions sampled by our spectra.  Therefore,  our results demonstrate that
the highest surface brightness regions (in this case the SSC) can give
a misleading  impression of the properties of the
YSP component associated with the diffuse light in the galaxy as a whole.

\subsection{The past and future of star formation in PKS1345+12}

Combining both imaging and spectroscopic techniques, we find evidence
for two distinct phases of recent star formation activity in
PKS1345+12: one occuring less than 6~Myr ago that is associated with
the  SSCs and perhaps also linked to current the AGN and ULIRG
activity;  the other occuring more than 40~Myr ago and associated with
the diffuse light.  This is consistent with previous HST imaging
studies which have shown evidence for more than one episode of star
formation activity in other merging systems including  the Antennae
\citep{Whitmore99}, NGC7252 \citep{Miller97,Maraston01}, and Arp220
\citep{Wilson06}.  Therefore our  results fit in with the emerging
trend that the star formation in merging systems is complex and
multi-modal. 

At this stage it is important to add the caveat that, although we have
assumed that all of the YSP in PKS1345+12 formed in starbursts
triggered by the merger,  an alternative explanation is that the older
YSP component associated with the diffuse light represents the
captured disk of one of the merging galaxies i.e.  the diffuse light
YSP is not directly associated with the merging process. To test this
latter scenario we have attempted to model the spectra of the diffuse
light for all apertures using template spectra for disk galaxies of Sa
and Sb morphological types taken from \cite{Kinney96}. We find good
fits with either Sa or Sb templates for Ap 1, Ap2, and Ap A and Ap
B. Although no good fits are found ($\chi^2_{red} > 2$) for the other
apertures, the \cite{Kinney96} templates represent averages for several
galaxies of the same morphological type, and each galaxy type template
encompasses a range of individual galaxy spectra, some of which
deviate substantially from the average SED.  Therefore we cannot
absolutely rule out the idea that the YSP detected in the diffuse
light for these apertures are associated with the disrupted disk(s) of
one or more of the merging galaxies. Overall it is entirely plausible
that the YSP associated with the diffuse light represent the captured
stellar population from one or more merging disk galaxies, rather than
stars formed in the merger itself

Are our results consistent with the numerical simulations of major
galaxy mergers? For the case of a merger between two galaxies with
substantial  bulges --- the likely scenario for PKS1345+12 --- the
simulations predict that the major merger-induced starburst occurs as
the two nuclei finally merge together, but that there will also be a
lower level of interaction-induced star formation activity at an
earlier stage due to the tidal effects of the interaction on the disks
of the individual merging galaxies \citep{Barnes96,Mihos96}.
Therefore it is plausible that the younger YSP represented by the SSC
and the ULIRG activity are all related to the high  gas densities
produced in the final stages of the merger as the nuclei merge
together. On the other hand, assuming that they are not captured disk
populations, the older YSP associated with the diffuse light may
represent the earlier phase of lower level interaction-induced star
formation. In this case,  the models predict that we should expect an
even higher level of star formation and AGN activity over the
$\sim10^7$~yr it will take for the nuclei to coalesce (\cite{Evans99}).

\subsection{Order of events in triggering the AGN}

Several recent studies of YSP have been used to estimate the
timescales and order of events for triggering the activity in radio
galaxies.  For their sample of three radio galaxies,
\cite{Tadhunter05} found relatively old post-starburst ages ($\sim$0.3
- 2.5Gyr) compared with the lifetimes of the radio sources.
\cite{Emonts06} also found  a significant ($\sim$0.3~Gyr) delay
between the starburst event and Radio-AGN activity for the radio
galaxy B2~0648+27. These results are consistent with a scenario in
which the radio activity is triggered relatively late in the merger
sequence, following the major merger-induced starburst. However,
\cite{Tadhunter05} also  discussed the possibility that some radio
sources have undergone multiple epochs of activity. Indeed,
\cite{Stanghellini05} detected low surface brightness extended radio
emission surrounding several objects in their sample of otherwise
compact ($D < 1$~kpc) radio sources, including PKS1345+12. Moreover,
this galaxy shows some evidence for continuity between the compact and
extended emission, which, if confirmed, would suggest that PKS1345+12
has had AGN-Radio activity over a longer timescale than the estimated
age of the GPS source ($<$0.1Myr).  Given the evidence for both
multiple epochs of star formation, including some YSP with ages less
than 6~Myr, and multiple epochs of radio jet  activity, we conclude
that there is no clear evidence for a delay between the merger-induced
starburst and the start of the AGN phase in PKS1345+12.

\subsection{SSC kinematics}

In Section 3.3 we presented evidence that the  SSC and their
associated HII regions are moving at high radial velocities relative
to the diffuse ISM detected (in projection) at  similar locations in
the halo of the host galaxy, despite the morphological evidence for
associations with diffuse ISM components.  It is interesting to
consider how such rapidly moving star clusters might form.

The first possibility recognises the fact that major galaxy mergers
are violent events, and considerable dissipation must take place
before the gas finally settles into a stable configuration in the
merged system. At all stages in a merger, but particularly the final
stages before the nuclei of the two merging galaxies coalesce, it is
likely that rapidly moving gas streams/tidal tails will rain down on
the central regions of the host galaxy. As individual gas streams
collide with the ambient gas or other gas streams, they will generate
shocks that may trigger star formation \citep{Barnes04}. The ionizing
photons generated in the shocks as the tidal streams interact with the
ambient gas may also lead to a local enhancement in the emission line
luminosity of the ambient gas, perhaps explaining the effect seen in
Figure 11. Therefore the formation of the SSC detected in PKS1345+12
may be associated with the overall settling/dissipation process that
must accompany the evolution of the ISM following a major galaxy
merger; it is possible that we are witnessing the events that will
eventually lead to the formation of a high velocity dispersion
globular cluster systems (e.g.\cite{Ashman92,Zepf93}) as the host galaxy
settles down to its final state as a quiescent elliptical
galaxy.

A second possibility is suggested by the detection of diffuse radio
emission extending $\sim$35 arcseconds ($\sim$83~kpc) to the north,
and $\sim$25 arecseconds ($\sim$60~kpc) to the  south, of the nucleus
of PKS1345+12, and encompassing the region of the SSC
\citep{Stanghellini05}. As discussed in the previous section, although
the major, high surface brightness radio activity is now taking place
on a sub-kpc scale close to the western nucleus
\citep{Stanghellini05}, the presence of diffuse radio emission on a
much larger scale provides evidence that there was an earlier phase of
activity in which the relativistic plasma jets escaped into the halo
of the merging system. In this phase, the radially expanding jets and
laterally expanding jet cocoon would have interacted strongly with the
warm/cool ISM associated with the merger, shocking the gas and
possibly triggering star formation \citep{Rees89,Begelman89}.
Because the shocked gas is accelerated to high velocities by the
jet-cloud interactions, it would be expected that any SSC formed in
such interactions would be moving at a high velocity relative to the
ambient ISM, as observed in PKS1345+12. Moreover, interaction with the
expanding cocoon could produce both positive and negative radial
velocities, depending on the interaction geometry. Therefore, the fact
that C1/C2 is blueshifted, while C4 is redshifted, is not a problem
for this model.

Note that the importance of jet-induced star formation as a mechanism
remains controversial. The direct observational evidence for the
mechanism remains sparse, confined to a few objects observed across a
large range of redshifts \citep{Best97,Dey97,Croft06}. There is  also
a recognition that hydrodynamic effects of the jet-cloud interactions,
as well as the heating effects of the accompanying quasar activity,
can be as destructive as they are conducive to star
formation. However, the recent detection of high velocity neutral
outflows against the cores of several compact radio sources
(\citealt{Morganti05a}a, \citealt{Morganti05b}b) ---  including
PKS1345+12 itself ---  demonstrates that, despite being accelerated to
high velocities, the gas behind a shock can at least cool to a neutral
phase. This lends weight to the plausibility of jet-induced star
formation as a mechanism.

\section{Conclusion}

We have reported a detailed investigation of the YSP in the ULIRG
PKS1345+12, combining information from both spectroscopic and imaging
methods. Our key conclusions can be summarised as follows:
\begin{itemize}
\item {\bf Reddening.} We find that the three SSC with good
photometric information in our study are all significantly
reddened. Clearly it is important to  have sufficient photometric
information with wide spectral coverage to remove any age-reddening
degeneracy and determine  accurate YSP properties, even for clusters
in the extended halo of a galaxy.
\item {\bf High resolution imaging versus spectroscopy.} This study
provides a clear demonstration of the fact that high resolution HST
imaging studies tend to be biased towards the youngest and/or most
massive individual star forming  clusters, and may give a misleading
impression of the star formation  histories of the dominant YSP in
terms of total mass (sampled by  the diffuse light in PKS1345+12).
\item {\bf The star formation history of PKS1345+12.} We find evidence
for a complex star formation history of the YSP in PKS1345+12, with at
least two major phases of star formation, one of which may be
associated with star formation in the merging disk galaxies at
an earlier stage of the merger.
\item {\bf Order of events.} Unlike some other radio galaxies, we
find no clear evidence in PKS1345+12 for a time delay between the
major merger-induced starburst and the AGN activity.
\item {\bf Star cluster kinematics.} On the basis of our analysis of
the  emission line kinematics, we deduce that the SSC are moving at
high velocity relative to ambient gas and the rest frame of the host
galaxy. The extreme SSC kinematics may be related to the process of
star formation as  infalling gas settles in the nuclear regions in the
final stages of the merger,  or to star formation triggered by the
expanding radio source as it drives shocks into the ISM.
\end{itemize}

\section*{Acknowledgments} 

We thank Enrique Perez and Simon Goodwin for useful  discussions, and
Katherine Inskip for technical IDL support. We also thank the
anonymous referee for useful comments that have helped to improve the
manuscript. JR and JH acknowledge financial support from PPARC. Based
on observations with the NASA/ESA {\it Hubble Space Telescope}, which
is operated by the Association for Research in Astronomy (AURA), Inc,
under NASA contract NAS5-26555. The William Herschel Telescope is
operated on the island of La Palma by the Isaac Newton Group in the
Spanish Observatorio del Roque de los Muchachos of the Instituto de
Astrofisica de Canarias.

\bibliographystyle{mn2e}
\bibliography{javi}

\end{document}